\documentclass[letterpaper]{natureprintstyle}
\usepackage{amsfonts}

\usepackage{eurosym}
\usepackage{graphicx,epsfig}
\usepackage{amsmath,amssymb,bbm}
\usepackage{color}
\usepackage[usenames,dvipsnames]{xcolor}
\usepackage[colorlinks=true,linkcolor=Red,citecolor=Green,linktoc=page]{hyperref}
\usepackage{multirow}
\usepackage[varg]{txfonts}
\usepackage{ulem}
\usepackage{fancyhdr}
\usepackage{mathtools}
\usepackage[caption=false]{subfig}
\usepackage{balance}
\usepackage{float}
\usepackage{flushend}

\def\subinrm#1{\sb{\rm#1}}
{\catcode`\_=13 \global\let_=\subinrm}
\mathcode`\_="8000
\def\upsubscripts{\catcode`\_=12 }

\upsubscripts

\begin{document}

\title{Controllable skyrmion chirality in ferroelectrics}
\author{Yu.\,Tikhonov,$^{1}$ S.\,Kondovych,$^{2,3}$ J.\,Mangeri,$^{4,5}$  M.\,Pavlenko,$^{1}$ 
L.\,Baudry,$^6$ A.\,Sen\'{e},$^2$ A.\,Galda,${^7}$  S.\,Nakhmanson,$^{8,5}$ O.\,Heinonen,$^{9}$ A.\,Razumnaya,$^{1}$ I.\,Luk'yanchuk,$^{2,10}$ \& V.\,M.\,Vinokur$^{9,11}$ }
\maketitle

\thispagestyle{fancy} 
\lfoot{\parbox{\textwidth}{ \vspace{0.2cm}
 \rule{\textwidth}{0.2pt}
\hspace{-0.2cm} \textsf{\scalefont{0.60}
   $^1$Faculty of Physics, Southern Federal University, 5 Zorge str., 344090 Rostov-on-Don,
Russia;
   $^2$University of Picardie, Laboratory of Condensed Matter Physics, Amiens,
80039, France;
    $^3$Life Chemicals Inc., Murmanska st. 5, Kyiv, 02660, Ukraine;
    $^4$Institute of Physics, Academy of Sciences of the Czech Republic, Na Slovance 2, 18221 Praha 8, Czech Republic;
    $^5$Department of Physics, University of Connecticut, Storrs, CT, USA;
    $^6$Institute
 				of Electronics, Microelectronics and Nanotechnology (IEMN)-DHS
 				D\'epartment, UMR CNRS 8520, Universit\'e des Sciences et
 				Technologies de Lille, 59652 Villeneuve d'Ascq Cedex, France;
 				$^7$James Franck Institute, University of Chicago, Chicago, Illinois 60637, USA;
 	$^8$Department of Materials Science \& Engineering and Institute of Material Science, University of Connecticut, Storrs, Connecticut 06269, USA;
 	$^9$Materials Science Division, Argonne National Laboratory,
    9700 S. Cass Avenue, Argonne, Illinois 60637, USA;
    $^{10}$Landau Institute for Theoretical Physics, Akademika Semenova av., 1A9 Chernogolovka, 142432, Russia;
    $^{11}$Consortium for Advanced Science and Engineering (CASE) University of Chicago, 5801 S Ellis Ave, Chicago, IL 60637, USA.
} \vspace{-0.2cm}
\begin{center}{\scalefont{0.87} \thepage}\end{center}}} \cfoot{}

\textbf{
Chirality, an intrinsic handedness, 
 is one of the most intriguing fundamental phenomena in nature\,\cite{Kelvin1894,Hegstrom1990, Wagniere2008, Berger2019}.
Materials composed of  chiral molecules find broad applications in areas ranging from nonlinear optics and spintronics to biology and pharmaceuticals\,\cite{Neufeld2018,Naaman2015,Yoo2019,Inaki2016}. 
However, chirality is usually an invariable inherent property of a given  material 
that cannot be easily changed at will.
Here, we demonstrate that ferroelectric nanodots support skyrmions the chirality of which can be controlled and switched. 
We devise protocols for realizing control and efficient manipulations of the different types of skyrmions.
Our findings open the route for controlled chirality with potential applications in ferroelectric-based information technologies.}

Chiral topological textures set the stage for a new generation of the chiral materials, where the chirality is extended over nano- and micro-scales. Nonuniform chiral states, helical, blue, and twist grain boundary (TGB) phases have been observed in cholesteric liquid crystals\,\cite{Bahr2001,Chandrasekhar2006}. 
Skyrmions, which are the chiral texture of a vector order parameter, such as magnetization or polarization density, have been attracting considerable attention in magnetic materials during the past decade\,\cite{Seki2016,Liu2016,Zhang2018} due to their potential applications in information technologies. However, a salient feature of these materials is the specific non-chiral symmetry, carried either by non-mirror-symmetric molecules in cholesterics or the antisymmetric spin exchange in magnetic systems 
 leading to the Dzyaloshinskii-Moriya spin interaction. 
 Recently, 
 an extension of the class of magnets hosting skyrmions 
 onto  systems 
 without Dzyaloshinskii-Moriya spin interaction has been reported\,\cite{Phatak2016,Navas2019}. However, the possibility of tuning the chirality of skyrmions in these systems remains an open question.

Although a pre-defined chiral symmetry is absent in ferroelectric materials, they were recently found to host a wealth of chiral topological excitations, including Bloch domain walls\,\cite{Tagantsev2001,Lee2009,Wojdel2014,Cherifi2017}, coreless vortices with a skyrmion structure\,\cite{baudry2012a,baudry2012b,baudry2014}, single skyrmions\,\cite{nahas2015,gonccalves2019},  skyrmion lattices\,\cite{das2019}, and Hopfions\,\cite{lukyanchuk2019}. 
A distinct feature of ferroelectrics is that the chirality appears as a result of the spontaneous symmetry breaking due to specific interplay of confinement and depolarization effects when the depolarization charges $\rho =\nabla\cdot\mathbf{P}$ rearrange to reduce their interaction energy, leading to the chiral twisting of the polarization. Importantly, the different chiralities (``left" and ``right" states) are energetically degenerate and hence inter-switchable.
However, performing such chirality-switching poses a challenge because of the non-chiral nature of the fundamental fields that could serve as a control parameter. 
We find that ferroelectric nanodots can provide rich phase diagrams as depolarization effects lead to an abundance of topological excitations, and
we demonstrate that ferroelectric nanodots harbor polarization skyrmions. 
In particular, we devise a system in which controlled switching between the opposite chiralities may be implemented by the applied electric field.

%%%%%%%%%%%%%%%%%%%%%%%%%%%%%%%%%%%%%%%%%%%%%%%%%%%%%%%%%%%%%%%%%%%%%%%%%%%%%%%%%%%%%%%%%
\begin{figure}[!b]
\center
\includegraphics[width=0.49\textwidth] {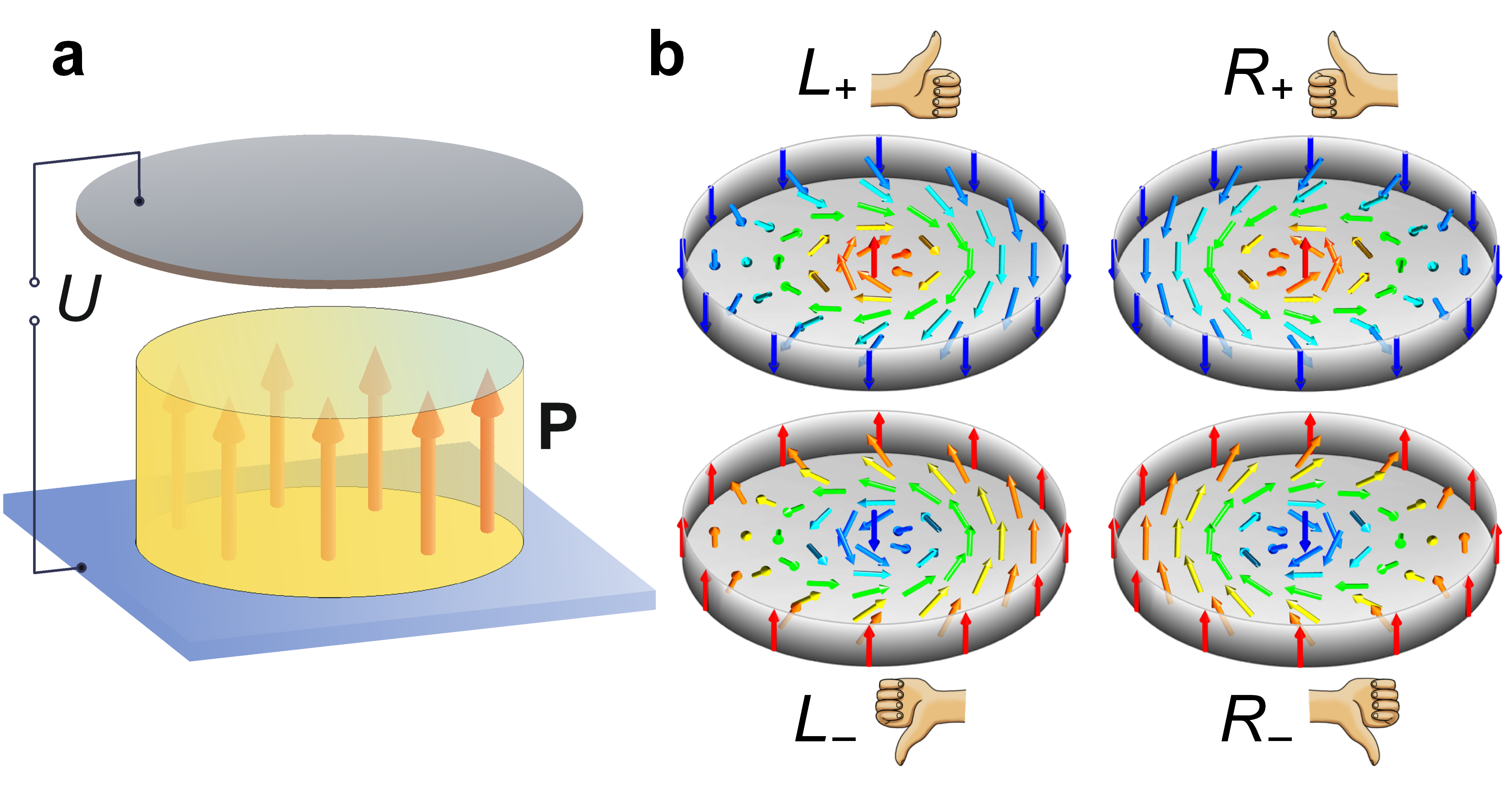} 
\caption{\textbf{Ferroelectric switch circuit and skyrmion states.} 
\textbf{a,}\,The circuit is controlled by
the external switching voltage $U$. The top electrode is separated  from the ferroelectric nanodot carrying the polarization topological states.
\textbf{b,}\,Four types of the skyrmions, differing by their chirality and polarity. The  hand pictograms define the classification of the skyrmions.
}
\label{FigSwitch}
\end{figure}
%%%%%%%%%%%%%%%%%%%%%%%%%%%%%%%%%%%%%%%%%%%%%%%%%%%%%%%%%%%%%%%%%%%%%%%%%%%%%%%%%%%%%%%%%%

Our target system is a ferroelectric nanodot in a shape of the disk deposited on a substrate (see Fig.~\ref{FigSwitch}a). We choose the lead titanate
pseudo-cubic perovskite oxide, PbTiO$_3$, as the model ferroelectric material. 
The typical nanodots that we use for observation of the skyrmion have diameter about 40\,nm and thickness about 20\,nm.
For calculations we employ the phase field approach, described in details in the Methods section.
The ferroelectric nanodot is located in a capacitor, the upper plate of which is separated from the sample either by vacuum or by a dielectric material with low dielectric constant. 
The thin lower electrode forms an interface between the nanodot and the substrate, which induces a weak strain caused by the ferroelectric lattice mismatch with the substrate. The strain results in an  
out-of-plain polarization anisotropy (see Methods for relevant parameters).
This system enables the creation and manipulation of 
topological nonuniform textures confined in the nanodot by applying voltage $U$ to the electrodes.

%%%%%%%%%%%%%%%%%%%%%%%%%%%%%%%%%%%%%%%%%%%%%%%%%%%%%%%%%%%%%%%%%%%%%%%%%%%%%%%%%%%%%%%%%
\begin{figure*}[!t]
\center
\includegraphics[width=14cm] {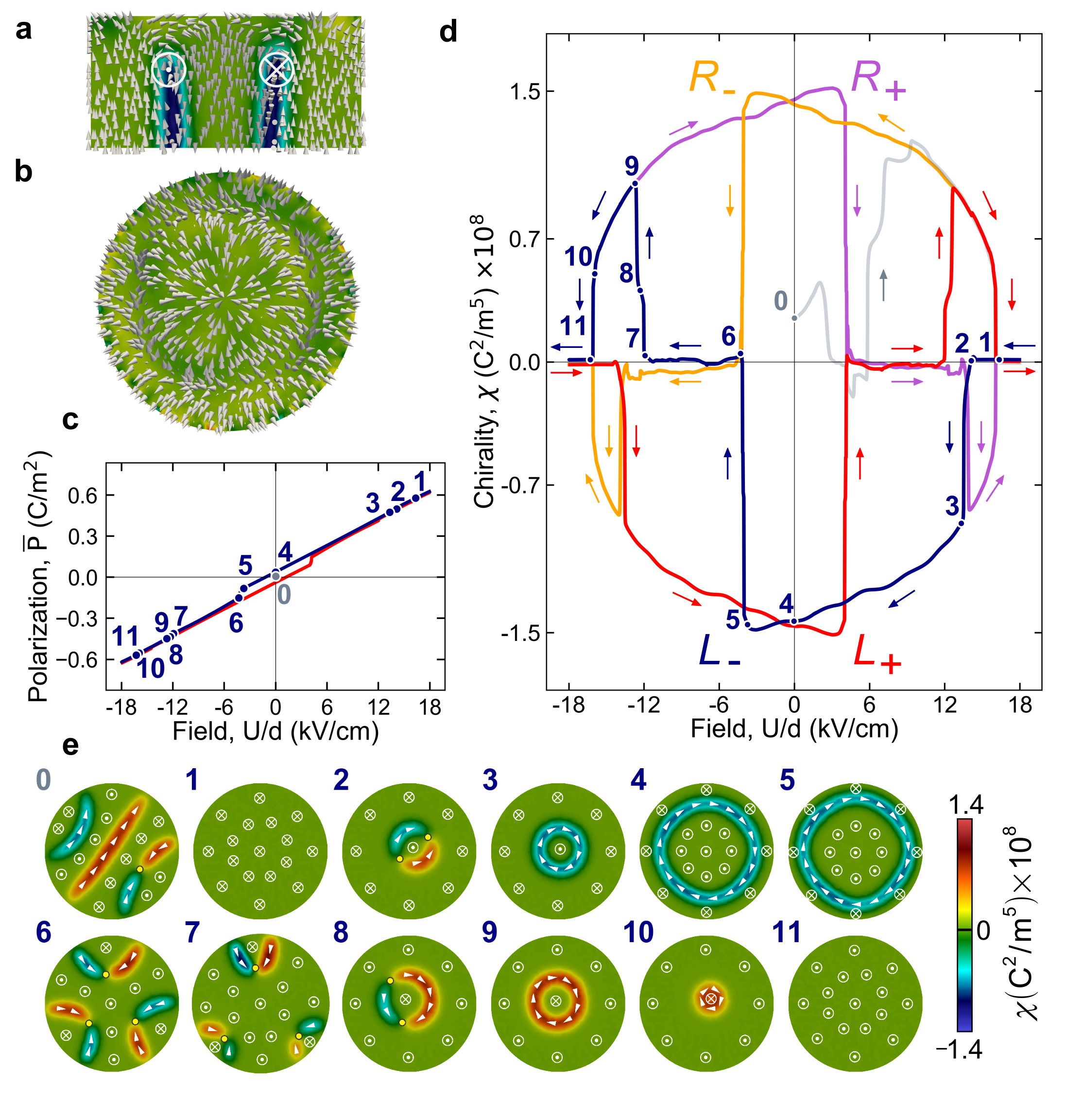} 
\caption{\textbf{Field-tuned topological states in the nanodot.} 
\textbf{a,}\,Cross section of the nanodot displaying the polarization distribution, with white arrows showing the direction of the polarization. The polarization rotation over 180$^\circ$ in the plane of the Bloch domain walls results in the chirality distribution, $\chi(r)$, shown by the colour map. The legend to the map is given below in low-right corner of the figure. The crossed circle, $\otimes$, at the domain wall denotes the polarization vector going into the cross-section plane, and the circle with the central dot, $\odot$, stands for the out–-of-–plane polarization.
\textbf{b,}\,The top view of the sinc-like distribution of the polarization at the near-surface layer of the nanodot.
\textbf{c,}\,Hysteresis behaviour of the polarization of the nanodot as a function of the applied field. The blue and red branches correspond to the up-down and down-up sweeps of the applied field. The numbers mark the different topological states of the polarization.
\textbf{d,}\,Hysteresis protocols of the chirality switching that allow to come to the $L_\pm$ and $R_\pm$ skyrmion states. The arrows show the direction of the sweep. 
The gray branch corresponds to the virgin curve of the poling of the nanodot. The blue and red branches again correspond to the up-down and down-up sweeps of the applied field. The violet and yellow branches correspond to the reversal of the field sweep from the blue and red branches, respectively. 
\textbf{e,}\,The distribution of the polarization and chirality in the original polarization state (0) and in the sequence of the topological states arising during the re-polarization of the nanodot by the applied field (view from the bottom) that follows the blue branch of panels \textbf{c} and \textbf{d}.  The yellow points mark the cross-section of the Bloch lines piercing the nanodot. 
}
\label{FigHyster}
\end{figure*}
%%%%%%%%%%%%%%%%%%%%%%%%%%%%%%%%%%%%%%%%%%%%%%%%%%%%%%%%%%%%%%%%%%%%%%%%%%%%%%%%%%%%%%%%%%

The emergence of these topological excitations is related to the effect of depolarization charges that arise because of the abrupt termination of the spontaneous polarization at the nanodot  surface. Induced depolarization fields  destabilize the uniformly-polarized state, resulting in 
a polarization texture corresponding to a self-consistent local energy minimum. In thin-film geometry, the depolarization effect yields Kittel domains\,\cite{bratkovsky2000}, structured as soft polarization stripes\,\cite{stephanovich2005,DeGuerville2005} (or parallel-anti-parallel vortices\,\cite{zubko2010,yadav2016}), or bubble domains in a skyrmion lattice structure\,\cite{das2019}. 
Confinement of these textures to within a nanodot gives rise to the competition 
between striped and cylindrical domains\,\cite{martelli2015}. 
We demonstrate that the latter configuration generates chiral skyrmions, provided that the substrate-induced uniaxial anisotropy is weak and that the polarization draws up a Bloch twisting of the polarization inside the domain wall. The left and right chiral skyrmions, shown in Fig.\,\ref{FigSwitch}b, are mnemonically visualized by left and right hands, where thumbs point in the direction of the polarization orientation in the core, 
which is referred to as skyrmion polarity, and the fingers curl along the orientations of the winding of the polarization around the skyrmion core 
corresponding to the skyrmion chirality.  Accordingly, we introduce the notations $L_+$, $L_-$, $R_+$, and $R_-$, where the "$L$" and "$R$" correspond to the left and right-hand skyrmions and "+" and "$-$" subscript denote the "up" and "down" polarities, respectively.   

The polarization distribution inside the skyrmion confined within the nanodot is shown in Fig.\,\ref{FigHyster}a.
The polarization texture preserves the structure of the polarization rotation inside the chiral 
Bloch-like circular domain wall
over almost entire height of the sample. It is within these chiral domain walls where the chirality of the nanodot is concentrated. 
At the top near-surface layer the polarization configuration assumes the sinc-like shape to form a N\'{e}el-type non-chiral 
skyrmion to maintain the polarization tangential to the surface, see the top view in Fig.\,\ref{FigHyster}b, in order to prevent the formation of surface depolarization charges.
The entire texture resembles the structure of the bubble domain in a double-periodic domain structure observed in ferroelectric superlattices\,\cite{das2019}, and has a topology of the Hopf fibration similar to that in the ferroelectric nanoparticles\,\cite{lukyanchuk2019}.
 In what follows we will construct protocols that enable formation of a skyrmion as well as electric field-tuned transitions between different polarization configurations.  
The switching between the different states will be described as 
switching between different mean chiralities, which we define as 
\begin{equation}
\bar{\chi}={1\over V}\int_{V}\,\chi(r)\,d\mathbf{r}\,, \qquad \chi=\mathbf{P}\cdot\left[\nabla\times\mathbf{P}\right],
\label{index}
\end{equation}
where $\chi(r)$ is a chirality density, and integration is performed over the nanodot volume $V$, so that the $\bar{\chi}>0$ for right- and $\bar{\chi}<0$ for left skyrmions. 
It is important to clearly distinguish between chirality defined by Eq.\,(\ref{index}) and 
identifying the objects that cannot be mapped to their mirror images by rotations and translations\,\cite{Kelvin1894},
and another swirling characteristics of the vector fields, vorticity, quantified as $\nabla\times\mathbf{P}$ and toroidal moment $\int \mathbf{r}\times(\mathbf{P}-\mathbf{\bar{P}})\,dV$.
The controlled manipulation by four distinct skyrmion states, $R_\pm$ and $L_\pm$, offers  the opportunity for implementing a platform for ferroelectric-based multivalued logic units\,\cite{martelli2015,Baudry2017a}.

In order to elaborate on the protocols for controlling and tailoring skyrmions, we investigate the response of the polarization in the nanodot to applied electric field $U$.  
The mean polarization ($\bar{P}(U)$) and the chirality ($\bar{\chi}(U)$) hysteresis curves are shown in Fig.\,\ref{FigHyster},  \textbf{c} and \textbf{d} respectively. The corresponding stages of the process are displayed in Fig.\,\ref{FigHyster}e. 
We first set the zero-field ferroelectric state at room temperature by quenching it from the paraelectric state with randomly oriented small-amplitude polarization. 
The resulting state, which we denote as 0-state, has in Fig.\,\ref{FigHyster}e a structure of four-band stable domain stripes\,\cite{martelli2015}. The domains are separated by the Bloch domain walls in which the direction of the polarization rotations determines the domain wall chirality that can be either positive (shown in red) or negative (shown in blue). 
At the loci where the chiralities of the opposite sign meet, a linear topological defect, so-called Bloch line (denoted as yellow dot in the cross-section images) penetrating the nanodot  forms.  The Bloch lines were observed in ferromagnetic domain walls\,\cite{Malozemoff2016} and predicted to appear in ferroelectrics\,\cite{Salje2014}. It is the dynamics of the Bloch lines that eventually controls the chirality switch in the nanodot. 
The precise distribution of the chiralities in the Bloch domain walls in the initial state of course depends on which random paraelectric configuration was quenched, but does not affect the polarization evolution after the initial poling in a large electric field.

To create and manipulate the skyrmion we ramp up the electric field by applying a voltage $U$ to the circuit, necessary to uniformly polarize the sample.  
The virgin curve (gray color line) passes through the polarization stripe states differing by the structure of domain walls which move and interswitch their chiralities,.
The evolution concludes with the jump into the skyrmion state at the latest stage before complete poling is achieved (state 1 in Fig.\,\ref{FigHyster}d.). 
We then start reducing the field strength to zero and, subsequently, having reversed the field direction, increase its magnitude. By analyzing the evolution of the polarization, we establish a protocol that allows us to switch the chirality of the system, as shown in Fig.\,\ref{FigHyster}d and e. 
We consider first the evolution of the system under decreasing field from state 1, depicted by the blue line in Fig.\,\ref{FigHyster}d. 
Upon decreasing the electric field below the threshold value, the polarization in the central region of the nanodot switches its orientation to the opposite one. As a result, the bubble domain (state 2) forms, for which the cylindrical domain wall partitions into two half-cylindrical segments with opposite chiralities separated by Bloch lines, the total chirality remaining zero.  
As the field strength is further reduced, the skyrmion $L_-$  (state 3) with negative chirality $\bar{\chi}$ forms as a result of merging the Bloch lines and a concomitant collapse of the positive-chirality segment, and a skyrmion appears (state 3). 
Further reducing the field strength, the thickness of the skyrmion core grows, leading to repartitioning of the up- and down-oriented polarization regions (state 4) and, hence in the $\bar{P}({U})$-dependence shown in Fig.\,\ref{FigHyster}c. 

The skyrmion texture remains as 
the field is completely removed (state 4), and even
as the field direction is reversed, although the texture now becomes metastable. At some negative field (state 5) the skyrmion becomes unstable and decays into a multi-domain state (state 6) with zero chirality. 
 The arc-shape domain walls form three pairs of segments with opposite chirality
that connect to the nanodot sides. The two segments of each pair are separated by a Bloch line, and the total chirality of each pair is zero.  
As the field magnitude is further increased, one of the domain walls pairs is rearranged to favor the nucleation of a cylindrical domain at the surface of the nanodot (state 7) with the abrupt propagation into the interior of the sample.  Two other pairs of domain walls disappear from the sample. The cylindrical bubble domain with two opposite-chirality domain walls settles at the  the center of the nanodot (state 8)  and further transforms to the $R_+$ skyrmion with $\bar{\chi}>0$ and negative mean polarization (state 9) through the collapse of the negative-chirality domain wall. As the field magnitude continues to increase, the core of the skyrmion shrinks and the system arrives at state 10. Finally, the system jumps to the uniformly polarized state 11 with the negative polarization orientation. If, however, the field at point 7 is reversed with a decreasing magnitude to reach zero 
(the violet branch in Fig.~\ref{FigHyster}d), the system remains in the $R_+$ skyrmion state at $U=0$.  With the further field increase in the positive direction, the system repeats the sequence 3-11 but with the opposite polarization and chirality. In other words, the system finally returns into the uniform up-polarized state 1. 

One can make the system evolve having started with the negative fields and the polarization poled in the negative direction. In this case, the system will follow the branches denoted by the red-yellow traces in Fig.~\ref{FigHyster}d. 
The emerging hysteresis branches are symmetric to the blue-violet ones with respect to $U\to-U$ reversal. The red branch corresponds to the blue one and the yellow branch corresponds to the violet one.
The corresponding polarization states for the potential $U$ are obtained from their counterparts from the blue-violet branches at the potential $-U$ by the reversing the sign of the $P_z$ component of the polarization $\mathbf{P}$.  By the proper sweep protocol of the electric field, one arrives therefore to the $R_-$ and $L_+$ skyrmions at $U=0$. Therefore, one sees that by the  appropriate set of the protocol, one can obtain and switch between all of the four skyrmion states, $L_{\pm}$ and $R_{\pm}$ with different chirality and polarity orientations.

It is important to note that the applied electric field does not possess its own chirality. This implies that the direction of the switch is determined rather by underlying local fluctuations in the chirality of the material which then serve as the nucleation centers of the emergent skyrmions. We describe this effect by introducing the fluctuating chirality field $\Lambda(r)$, which enters the system energy functional as the additional term 
$-\Lambda\,(\mathbf{P}\cdot\nabla \times\,\mathbf{P})$.
In our numerical experiments, these spatial fluctuations are implemented via 
generating random tetrahedral configurations, maintaining the approximately constant mesh size. 
Altering a particular mesh changes the sign of emerging  1-2 and/or 6-8 jumps at the blue branch. The triggering effect of mesh fluctuations is verified by the mirror reflection of the discretization mesh leading to changing the sign of the chirality jump to the opposite one.
Remarkably, even small fluctuations in $\Lambda$ lead to switching chirality. This implies
an opportunity for laser-activated manipulation of the polarization\,\cite{Stoica2019} employing the circular  polarized irradiation of the optical tweezers for controlling the direction of the chirality switch.

%%%%%%%%Methods%%%%%%%%%%%%%%%%%%

\section*{METHODS}
~~\newline
\small{
The polarization of the strained nanodot is obtained from the
minimization of the free energy functional, depending on the polarization, $\mathbf{P}=(P_1,P_2,P_3)$, and the electrostatic potential, $\varphi$,
\begin{gather}
F=\int \Big(\left[ a_{\mathit{i}}^{\ast }(u_m,T)P_{\mathit{i}}^{2}+a_{\mathit{ij}}^{\ast }P_{%
\mathit{i}}^{2}P_{\mathit{j}}^{2}+a_{\mathit{ijk}}P_{\mathit{i}}^{2}P_{%
\mathit{j}}^{2}P_{\mathit{k}}^{2}\right] _{\mathit{i}\leq \mathit{j}\leq 
\mathit{k}}  \notag \\
+\frac{1}{2}G_{\mathit{ijkl}}(\partial _{\mathit{i}}P_{\mathit{j}})(\partial
_{\mathit{k}}P_{\mathit{l}})+\left( \partial _{\mathit{i}}\varphi \right) P_{%
\mathit{i}}-\frac{1}{2}\varepsilon _{0}\varepsilon _{b}\left( \nabla \varphi
\right) ^{2} -\Lambda(\mathbf{P}\cdot\nabla \times\,\mathbf{P})  \, \Big)%
\;d^{3}r\,, \label{Functional}
\end{gather}%
where the summation over the repeated indices $i,j,...=1,2,3$ (or $x,y,z$) is
performed. The numerical parameters are specified for the PbTiO$_3$ nanodot, strained by the substrate with the compressive misfit strain $u_m\simeq -0.002$.
The first square brackets term of (\ref{Functional}) stands for
the Ginzburg-Landau energy of the strained ferroelectric film\,\cite{pertsev1998}, written in the form given in\thinspace \cite%
{Baudry2017a}. The 2nd-order coefficients depend on the misfit strain $%
u_{m}$ and temperature $T$ and are expressed as 
$a_{1}^{\ast }=a_{2}^{\ast }=3.8\times 10^{5}(T-479\,^\circ C)-11\times 10^{9}\,u_{m}$\,C$^{-2}$m$^{2}$N$^{-1}$ and 
$a_{3}^{\ast }=3.8\times 10^{5}(T-479\,^\circ C)+9.5\times 10^{9}\,u_{m}$\,C$^{-2}$m$^{2}$N$^{-1}$. 
The strained renormalized 4th-order coefficients partially account for the elastic interactions, obey the tetragonal symmetry conditions and equal to 
$a_{11}^{\ast}=a_{22}^{\ast}\simeq 0.42\times 10^{9}$\,C$^{-4}$m$^{6}$N, 
$a_{33}^{\ast }\simeq 0.05\times 10^{9}$\,C$^{-4}$m$^{6}$N, 
$a_{13}^{\ast }=a_{23}^{\ast }\simeq 0.45\times 10^{9}$\,C$^{-4}$m$^{6}$N and  
$a_{12}^{\ast }\simeq 0.73\times 10^{9}$\,C$^{-4}$m$^{6}$N. 
The  6th-order coefficients conserve the cubic symmetry, 
$a_{111}=a_{222}=a_{333}\simeq 0.26\times 10^{9}$\,C$^{-6}$m$^{10}$N, 
$a_{112}=a_{113}=a_{223}\simeq0.61 \times 10^{9}$\,C$^{-6}$m$^{10}$N,  and
$a_{123}\simeq-3.7\times 10^{9}$\,C$^{-6}$m$^{10}$N.
The second term of\,(\ref{Functional}) corresponds to the gradient energy. The gradient energy coefficients $G_{ijkl}$ are obtained by the cubic symmetry permutations of the non-equivalent representatives  $G_{1111}=2.77\times 10^{-10}
$\thinspace C$^{-2}$m$^{4}$N, $G_{1122}=0$, and $G_{1212}=1.38\times 10^{-10}
$\thinspace C$^{-2}$m$^{4}$N\,\cite{Wang2004}.
The next two terms in\,(\ref{Functional}) correspond to the electrostatic energy, written in terms of the electrostatic potential $\varphi$\,\cite{landau1984}. Here, $\varepsilon _{0}=8.85\times 10^{-12}$\thinspace
C\thinspace V$^{-1}$m$^{-1}$ is the vacuum permittivity, and $\varepsilon
_{b}\simeq 10$ is the background dielectric constant of the non-polar ions\,\cite{Mangeri2017}. The last term of\,(\ref{Functional}) emulates the interaction of the ferroelectric polarization with the material chirality fluctuation, described by the parameter $\Lambda$. 
In most simulations, we took $\Lambda=0$. The local fluctuation, defining the direction of the chirality jumps  naturally arise  due to the random mesh configuration. To calibrate the effect of the mesh fluctuations we swept the value of $\Lambda$ and found that at 
 the threshold value $\Lambda_c\simeq 7\times 10^{-5}$\,C$^{-2}$m$^{3}$N, the direction of the chirality jump 2-3 at Fig.\,\ref{FigHyster}d changes to the opposite one.

The phase-field simulations were
performed using the FERRET package\,\cite{Mangeri2017}, designed
for the multi-physics simulation environment MOOSE\,\cite{Gaston2009}. We solved the time-dependent relaxation equation 
$-\gamma\,\partial P_{\mathit{i}}/\partial t=\delta F/\delta P_{\mathit{i}}$,
where $F$ is the total free energy functional including the electrostatic potential $\varphi$. The latter is found at each step of the relaxation as a solution of the Poisson equation 
$\varepsilon _{0}\varepsilon _{b}  \nabla ^{2}\varphi =\partial _{\mathit{i}}P_{\mathit{i}}$.
The relaxation parameter $\gamma $ was taken unity since
the time scale is irrelevant to the problem. The details of simulations are
similar to those described in\,\cite{Mangeri2017}. The geometry
and conceptual setup of the simulated system are shown in Fig.\,\ref{FigSwitch}\,a. We selected the diameter and the thickness of the nanodot as 40 and 20\,nm respectively as the optimal geometry for the skyrmion's observation.  
The driving field was controlled by the voltage $U$,
applied to the electrodes. The upper electrode was separated from the nanodot by the empty, or low-$\varepsilon$ dielectric space of the thickness of $60$\,nm. 
The initial paraelectric state with the small
randomly-distributed polarization was used as an initial condition for the quench to the original (virgin) state. Then, the quasi-static field variation protocols were used
 with the polarization distribution at the previous stage taken as the initial condition. 
The different finite-element meshes were used to ensure the stability of the process.
}

\medskip
\normalsize{
\noindent \textbf{References}
\vspace{-0.7cm}
}

\bibliography{skybib}

\smallskip
%\item [Acknowledgements]
\noindent \textbf{Acknowledgements.} This work was supported by 
the US Department of Energy, Office of Science, Basic Energy Sciences, Materials Sciences and Engineering Division (O.H. V.M.V and partly I.L); by
the H2020 RISE-ENGIMA, RISE-MELON and ITN-MANIC actions (Y.T., A.R. and I.L.), French goverment for Bourse Ostrogradski (Y.T.), and by the Southern Federal University, Russia (A.R. and Y.T.).

\end{document}